**Triplet Dimerization Crossover driven by Magnetic Frustration in $In_2VO_5$**

Simon A.J. Kimber[1,2], Mark A. de Vries[1,2], Javier Sanchez-Benitez[1,3],

Konstantin V. Kamenev[1,3], J. Paul Attfield[1,2*]

[1]*Centre for Science at Extreme Conditions, University of Edinburgh, Erskine Williamson Building, King's Buildings, Mayfield Road, Edinburgh EH9 3JZ, United Kingdom*

[2]*School of Chemistry, University of Edinburgh, Joseph Black Building, King's Buildings, West Mains Road, Edinburgh EH9 3JJ, United Kingdom*

[3]*School of Engineering & Electronics, University of Edinburgh, King's Buildings, Mayfield Road, Edinburgh EH9 3JL, United Kingdom*

$In_2VO_5$, containing magnetically frustrated zig-zag chains, shows a remarkable magnetic crossover at 120 K between paramagnetic states with positive (17 K) and negative (-70 K) Weiss temperatures. Magnetic moment and entropy data show that the $V^{4+}$ $S = 1/2$ spins condense into $S = 1$ triplet dimers below the crossover. A further freezing of the antiferromagnetically coupled triplet dimers into a global singlet state is observed at 2.5 K, with no long range magnetic order down to 0.42 K and in fields up to 9 T. No structural V-V dimerization is observed by high-resolution X-ray diffraction down to 10 K, but a subtle lattice anomaly evidences a spin-lattice coupling in the triplet dimer state. This is assigned to longitudinal oxygen displacement modes that reduce frustration within the chains and so couple to the spin dimer fluctuations.

The properties of low dimensional quantum magnets have been studied extensively, in particular the one-dimensional antiferromagnetic (AF) S = 1/2 chain, for which the susceptibility has been calculated to very high accuracy [1]. The addition of a next nearest neighbour, frustrating, AF exchange interaction ($J_2$) produces an array of exotic instabilities and ground states; this model is realised by the zig-zag chain shown in Fig. 1. The much studied case with $J_1$, $J_2 > 0$ (AF–AF) is unstable with respect to a singlet dimerised state with a spin gap when the ratio $\alpha = -J_2/J_1 > 0.2411$ [2,3], as exemplified by the spin-Peierls material $CuGeO_3$ [4]. The zig-zag chain with ferromagnetic $J_1$ and antiferromagnetic $J_2$ (F–AF) coupling has been less studied, however recent work has shed new light on the phase diagram and thermodynamics of this model. The ground state is reported to be ferromagnetic for $0 \leq \alpha \leq 1/4$ and an antiferromagnetic helix for $\alpha > 1/4$ [5-7], which has been realised in several materials containing S = 1/2 $Cu^{2+}$[8-10]. In this paper we report magnetic susceptibility, heat capacity and synchrotron powder x-ray diffraction measurements for $In_2VO_5$. Our results reveal a novel, spin dimerization instability in an F-AF S = 1/2 zig-zag chain. Below 120 K, $In_2VO_5$ behaves as an unfrustrated chain of antiferromagnetically coupled S =1 triplet dimers which show a further condensation into a global singlet ground state at 2.5 K. $In_2VO_5$ has an orthorhombic structure (space group *Pnma*) consisting of edge-sharing $VO_6$ chains along the *b* axis, separated by sheets of diamagnetic $InO_6$ octahedra [11]. Two recent theoretical papers have investigated magnetic exchange in $In_2VO_5$ and have reported contradictory results of frustrated antiferromagnetic [12] or ferromagnetic exchange interactions [13].

A high purity sample of $In_2VO_5$ was prepared by grinding small single crystals, grown by the previously reported method [11]. Our samples are black and semiconducting. The magnetic susceptibility of $In_2VO_5$, measured with a Quantum

Design MPMS system in a 150 Oe field, and the inverse susceptibility are shown in Fig. 2. Two distinct linear regions are seen in the inverse susceptibility with a crossover at 120 K. A Curie-Weiss fit in the range 140 – 300 K, gives a Weiss temperature of $\Theta$ = 17 K, and a moment of 1.81 $\mu_B$, consistent with an S = 1/2 system with g = 2.09. Below 120 K, the Weiss temperature changes sign ($\Theta$ = -70 K) and the paramagnetic moment increases to 2.2 $\mu_B$. A crossover between positive and negative $\Theta$ regimes is very unusual, and could signify a transition at which the structure and hence the exchange pathways are significantly altered, but this is ruled out by the diffraction results below. Our alternative explanation, which is supported by the heat capacity measurements, is that the zig-zag S = 1/2 chains in $In_2VO_5$ undergo a crossover into a chain of triplet dimers, as shown in Fig. 1. This is a plausible ground state for the F-AF zig-zag chain with $\alpha \sim 1$ [14]. The paramagnetic moment of 2.1 $\mu_B$ calculated for the triplet (S = 1) dimerized state, with g = 2.09, is in good agreement with the observed value of 2.2 $\mu_B$ below the 120 K crossover.

To determine whether $In_2VO_5$ undergoes a structural transition at 120 K, synchrotron powder x-ray diffraction data were collected at temperatures 10 – 280 K with wavelength $\lambda$ = 0.45621 Å, using the high resolution instrument ID31 at the ESRF, Grenoble. The data were fitted well by the previously reported *Pnma* structure, using the GSAS suite of programs [15,16], and no gross structural transition (e.g. V-V dimerization) or superstructure was seen down to 10 K. However, the lattice parameters show a small anomaly at 120 K, with a minimum in the *b* (chain) axis length and a subtle discontinuity in the cell volume (Fig 3a). This evidences a coupling of the local spin correlations to the lattice. However, no significant change to the atomic coordinates, and hence to the bond angles that influence the exchange pathways, are observed around the 120 K crossover. The lattice microstrain [17] parallel to *b*, $s_\parallel$[010], which quantifies local variations in the axis length, is larger than the perpendicular component and increases on cooling from 300 K but then decreases at low temperatures (Fig. 3b). This is consistent with a more highly correlated spin state, if spin-lattice coupling is significant.

The heat capacities ($C_p$) of $In_2VO_5$ and the diamagnetic analogue $In_2TiO_5$ were measured at 0.42 – 200 K using a Quantum Design PPMS system and are plotted in Fig 4. A $C_p/T$ peak is seen at 5 K for $In_2VO_5$ and a broad feature is also observed around 120 K. The magnetic heat capacity of $In_2VO_5$ was estimated by subtracting the $In_2TiO_5$ data and the resulting $C_p$(mag)/T and magnetic entropy, $S_{mag}$, are shown in Fig. 5. Two magnetic entropy releases are observed. On warming up to ~100 K, the entropy tends towards the 1/2Rln3 ≈ 4.5 J/mol.K value expected for 1/2 mole of S = 1 dimers. Above 120 K, an additional contribution increases the total magnetic entropy to approximately the Rln2 ≈ 5.8 J/mol.K value for the monomeric S = 1/2 state. Hence the magnetic heat capacity variation for $In_2VO_5$ confirms the triplet dimerization model proposed above.

A separate magnetic freezing transition below 5 K is evidenced in the low temperature magnetic susceptibility and heat capacity measurements. The broad magnetic susceptibility maximum at 2.5 K (Fig. 2, inset) shows that the global magnetic ground state of $In_2VO_5$ is a spin singlet, as expected for an antiferromagnetic S = 1 chain. The magnetic heat capacity also shows a shoulder around 4 K (Fig. 4, inset), which decreases slightly in magnitude and moves to higher temperature in a 9 T field. The breadth of this peak and its robustness to a 9 T field shows that it does not signify a long range magnetic ordering transition. (For long range order, a field estimated as $k_B T_{peak}/g\mu_B$ ≈ 4 T would be sufficient to suppress the peak). Furthermore, the heat capacity shows a power law scaling, $C_p$(mag) ~ $T^{1.89(2)}$ below 1 K. Linear scaling is expected for a truly one-dimensional system, but similar values of 1.7-1.8 are reported in several quasi-1D systems (see footnote 21 in ref. 10).

The S = 1 dimer chain is one of the ground states predicted by the F–AF alternating Heisenberg S = 1/2 chain model. The ground state varies smoothly from the dimerized (triplet) S = 1 Haldane chain to the dimerized (singlet) S = 0 chain as the ratio of exchange interactions changes [18,19]. There are few experimental realisations of the triplet chain, but IPACuCl$_3$ is a notable example, showing a susceptibility and excitations characteristic of an S = 1 chain with a Haldane gap, and field induced Bose-Einstein condensation [20-22]. Spontaneous triplet dimerization has not been reported in frustrated networks and so the unconventional ground state of In$_2$VO$_5$ is notable. Cu$^{2+}$ zig-zag chains with comparable exchange interaction ratios $\alpha$ = 0.25 – 3 instead form incommensurate spin ordered ground states [10].

We propose that the novel ground state found in In$_2$VO$_5$ evidences a high sensitivity of the $J_1$ interaction to small fluctuating lattice distortions. A weak spin-lattice coupling accompanying the dimer formation in Fig. 1 leads to inequivalent $J_1$ (intradimer) and $J_1$' (interdimer) interactions. Displacements that give $J_1/J_1$' > 1 reduce magnetic frustration and so the spin dimer fluctuations will couple to these modes. The small anomalies observed in the *b*-axis parameter and microstrain around the 120 K dimerization crossover are consistent with changes of longitudinal [010] phonon modes. The sensitivity of the $J_1$ exchange interactions to local distortion lies in the orientation of the magnetic orbitals in In$_2$VO$_5$. The unpaired electrons are localised in $d_{xy}$ type orbitals, perpendicular to the short V-O bonds (marked on Fig. 1), which are effectively orthogonal within the $J_1$ pathways in a symmetric zig-zag chain, leading to ferromagnetic superexchange. A small distortion such as displacement of oxygen atoms parallel to the chain direction removes the orthogonality and introduces a kinetic exchange component that rapidly decreases the strength of one of the now-inequivalent ferromagnetic interactions ($J_1'$). Hence, longitudinal oxygen vibrations can couple to variations in $J_1/J_1'$ and the corresponding triplet dimer fluctuations. The freezing of the S=1 dimers below 5 K may be accompanied by the ordering of local displacements, and further studies of the low temperature properties of $In_2VO_5$ will be worthwhile to discover whether a Haldane gap or lattice distortions are observed. High field investigation of possible Bose Einstein condensation, similar to that in $IPACuCl_3$ [22] also merits investigation.

In summary, $In_2VO_5$ shows a novel crossover between S = 1/2 and S = 1 dimer phases at 120 K. This instability reveals a new ground state for the frustrated zig-zag chain. Spin-lattice coupling is evidenced by subtle changes in the cell parameters and strains around the crossover, and this is ascribed to coupling between triplet dimer spin fluctuations and longitudinal oxygen vibrations within the zig-zag chains. The spin dimers freeze into a global singlet state below 5 K without apparent long range magnetic order.

S.A.J.K acknowledges the E.P.S.R.C. for funding and for provision of beam time, J.P.A acknowledges the Leverhulme trust for support. The authors acknowledge Dr. A.N. Fitch (ESRF) and Dr. J.W.G. Bos (UoE) for assistance with the synchrotron diffraction experiments and Prof. S.T. Bramwell for useful discussions.

[010] and the normal to the (hkl) diffraction plane. $s_i = 0.036\%$ is the instrumental broadening contribution found by fitting peaks from a standard silicon powder.

**Figure Captions:**

Figure 1: Zig-zag chain arrangement of $VO_6$ octahedra in $In_2VO_5$ showing the magnetic exchange interactions and the triplet dimers formed below 120 K. The short V-O bonds perpendicular to the chain direction are shown as solid lines.

Figure 2: Magnetic susceptibility of $In_2VO_5$ measured in a 150 Oe field, and the inverse susceptibility with Curie-Weiss shown in the ranges 30 - 100 K and 150 – 300 K and extrapolated to low temperatures. The inset shows the low temperature susceptibility maximum.

Figure 3: Temperature evolution of (a) the *b*-axis length and cell volume and (b) the lattice microstrains parallel and perpendicular to [010], from the high resolution synchrotron X-ray diffraction study of $In_2VO_5$.

Figure 4: Heat capacities of $In_2VO_5$ and $In_2TiO_5$ plotted as $C_p/T$ against temperature. The inset shows magnetic part of the heat capacity of $In_2VO_5$ measured in zero and 9 T fields with a power law fit to the zero field data below 1 K.

Figure 5: Magnetic heat capacity of $In_2VO_5$ (shown as $C_p(mag)/T$) and the integrated entropy versus temperature, showing the expected limits for triplet dimer (1/2Rln3) and S =1/2 monomer (Rln2) states.

1)

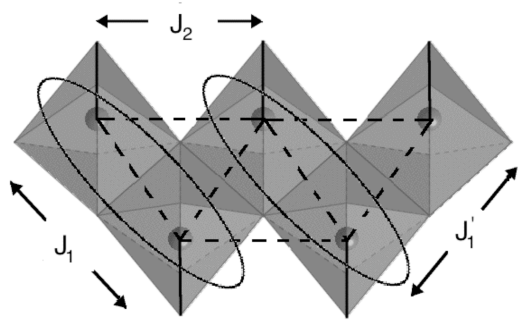

2)

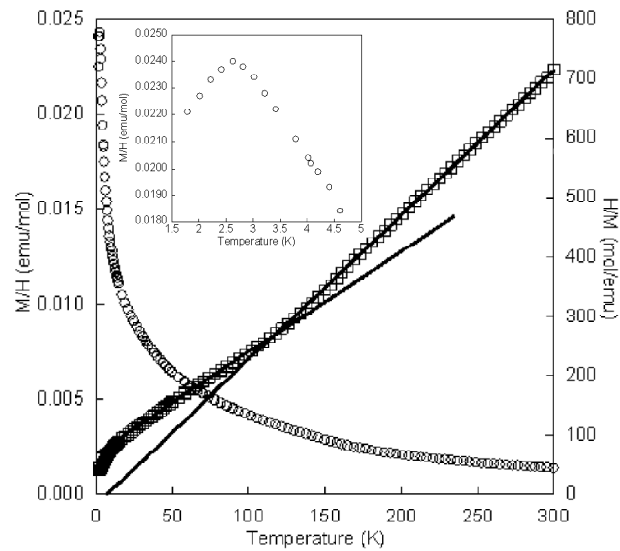

3)

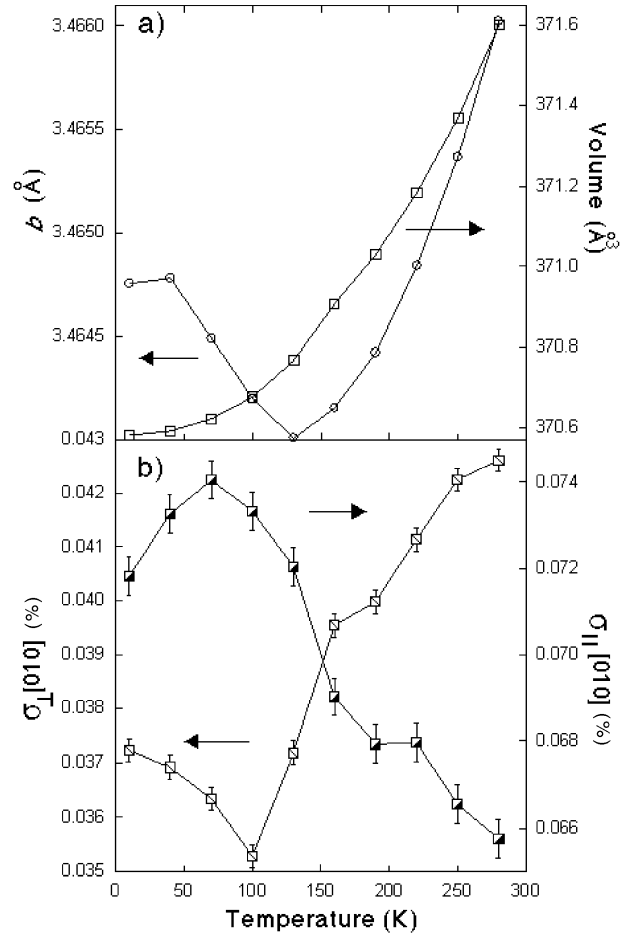

4)

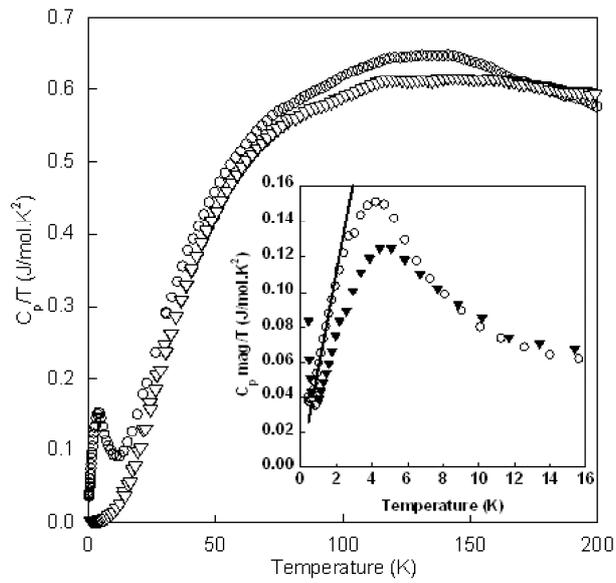

5)

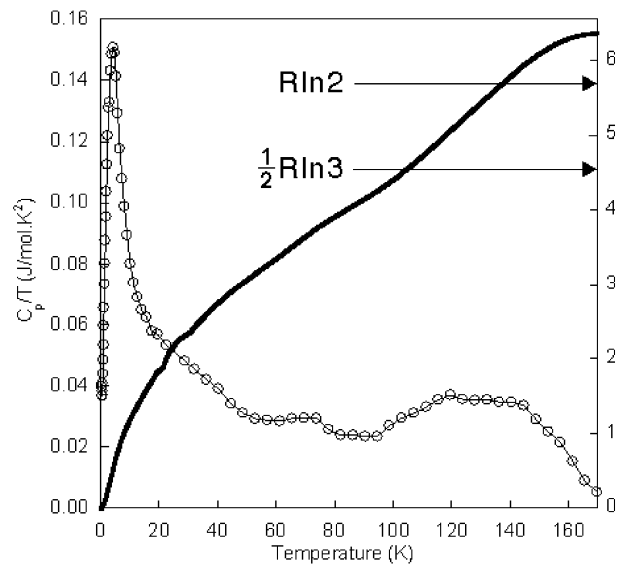